# Fast Timing With Silicon Carbide Low Gain Avalanche Detectors

P. Barletta[1], M. Cerullo[1], C. Haber[2*], S.E. Holland[2], J. Muth[1], B. Sekely[1]

1) North Carolina State University; 2) Lawrence Berkeley National Laboratory

*contact: chhaber@lbl.gov

**Abstract**

4H-Silicon Carbide, when considered as a material for the fabrication of Low Gain Avalanche Detectors for particle timing and position measurement, offers potential advantages over Silicon. We discuss an ongoing study of this material aimed at the fabrication and test of prototype fast timing sensors. This work is well aligned with technical directions identified in the recent Department of Energy study, "Basic Research Needs for High Energy Physics Detector Research and Development".

**Introduction**

In this contributed paper to the Snowmass proceedings we discuss ongoing work to study and develop Low Gain Avalanche Detectors (LGADs) using 4H Silicon Carbide (4H-SiC) rather than Silicon. This effort is very well aligned with all three of the Priority Research Directions (PRD's) identified for future tracking detectors in the recent DOE Basic Research Needs Study [1]. The PRD's are (PRD18) fast timing, (PRD19) new materials and processes, and (PRD20) low mass scalable tracking systems.

The work is being carried out as a collaboration between Physics Division staff at Lawrence Berkeley National Laboratory and large bandgap device and processing researchers at North Carolina State University (NCSU). NCSU, located in the Raleigh Durham Research Triangle Area, is a center for research and development on large bandgap devices, mainly for power applications. Many companies with expertise in large bandgap materials are located in this area as well.

Present upgrades to high energy and heavy ion collider detectors, and also proposed detectors at the new Electron Ion Collider, have embraced the importance of fast (~10 ps) timing on individual charged particles to enable 4D tracking, pileup rejection, and particle ID. The DOE BRN on Instrumentation states as a key goal "Develop high spatial resolution pixel detectors with high per-pixel time resolution to resolve individual interactions in high-collision-density environments". Much work on fast timing has centered on the Silicon LGAD concept [2]. This is an enhancement to the familiar position sensitive silicon detector (strips, pads, and pixels) which adds a moderate gain layer below the rectifying implant. However LGADs suffer from the same challenges as regular Silicon detectors as they are susceptible to bulk radiation damage and must, consequently, be operated a very low temperatures. This low temperature leads to bulky and complex cooling requirements.

The goal of this project is to explore the LGAD concept realized initially in Silicon Carbide (4H-SiC) rather than in Silicon. This material has already been studied and demonstrated as a radiation detector but in niche applications [3]. It has not found widespread application in large HEP or NP trackers, due, in part, to small signal and difficult fabrication. However, SiC detectors can, in principle, operate at very high temperatures, and are known to exhibit significant radiation hardness and extremely low leakage currents. Also, SiC has, in the past, lacked the industrial base of Silicon technology, making it hard to procure in significant quantities. Recent interest in batteries and power semiconductors for electric power switching has led to a significant increase in interest, research, and the availability of alternative semiconducting materials, including SiC [4]. Furthermore, recent advances in wafer size and quality, are enabling compatibility with processing tools developed for silicon devices and allowing the price per device to sharply decrease.

An LGAD, realized in SiC, overcomes the problem of small signal, and could therefore be a scalable, low mass solution to the needs for fast timing at future colliders. We would hope to benefit from the growing research and manufacturing base for SiC. This may be an optimal time to purse research in this area.

Work in this area is already underway in China and has been reported within the RD50 collaboration [5,6].

**4H-SiC versus Si**

Table 1 below summarizes and compares the properties of Silicon and 4H-SiC

| Property | Silicon | 4H-SiC |
|---|---|---|
| Bandgap (eV) | 1.12 | 3.27 |
| Energy per ion pair (eV) | 3.6 | 7.78 |
| Dielectric constant | 11.7 | 9.7 |
| Breakdown field (MV cm$^{-1}$) | 0.3 | 3 |
| Density (g cm$^{-3}$) | 2.3 | 3.2 |
| dE/dx minimum (MeV cm$^{-1}$) | 2.7 | 4.4 |
| Atomic number Z | 14 | ~10 |
| Electron mobility (cm$^2$ V$^{-1}$ s$^{-1}$) at 300K | 1300 | 800-1000 |
| Hole mobility | 460 | 115 |
| Saturated electron velocity (10$^7$ cm s$^{-1}$) | 1 | 2 |
| Threshold displacement energy (eV) | 13-20 | 22-35 |
| e-h pairs per micron | 80 | 57 |
| Thickness for equivalent signal ($\mu$m) | 1 | 1.57 |
| Thermal conductivity (W m$^{-1}$ K$^{-1}$) | 130 | 370 |
| Radiation length | 9.4 | 8.7 |
| Impact ionization coefficient | $\alpha_e > \alpha_h$ | $\alpha_e < \alpha_h$ |

Cells highlighted in brown point to key comparisons. 4H-SiC has a high breakdown field. The electron mobility in SiC is considerably higher than the hole mobility in Silicon. The primary ionization signal in SiC is about 70% of that in Silicon. The thermal conductivity of 4H-SiC is substantial. Finally the relative magnitude of the ionization coefficients in SiC are the opposite of those in Silicon [7]. Avalanches in Silicon must be initiated by electrons while in SiC it is by holes. This last property, combined with the higher electron mobility is a fortunate coincidence. Typically in Silicon the LGAD structure is made with a p-type drift region and gain region, and an n++ implant. In SiC the natural structure is therefore an n-type drift and gain region, and a p++ implant. This results, theoretically, in a faster timing performance for a 4H-SiC LGAD.

These device structures are compared in Figures 1 and 2.

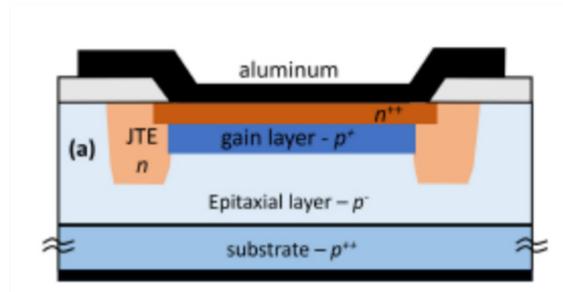

Figure 1: Idealized Silicon LGAD structure [8].  With a p-type drift region the device initially collects primary electrons and then the fast signal is due to gain holes drifting back across the p-epi region.

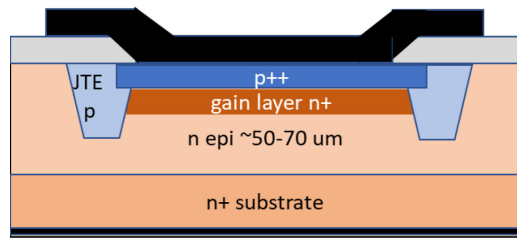

Figure 2: Idealized 4H-SiC LGAD structure.  With an n-type drift region the device collects primary holes and then the fast signal is due to gain electrons drifting back across the n-epi region.

**Results and Plans**

In order to understand the signal and operating conditions of a 4H-SiC LGAD we utilized an analytical 1D electrostatic model.  This model provided basic operating points and physical requirements for epitaxial thicknesses and doping levels.  We searched for a vendor(s) who could provide epitaxial wafers which met our requirements.  The low doping levels preferred for the n-epi layer are a challenge and typically lay outside, or at the limits of, what is normally available (note a similar situation existed for high resistivity Silicon in the early years of position sensitive Silicon detector development.)  We found, and contracted with, two vendors and have received wafers, built to meet our specifications, from one, to date.

We are also investigating more sophisticated 2D and LGAD signal simulations.

SiC is a more difficult material to process than Silicon.  In order for the power electronics community to realize the advantages of SiC it had to overcome a number of challenges.  SiC requires a high temperature (1300°C+) oxidation and post-ox anneal which is above the quartz softening temperature of ~1100°C.  Ion implantation also requires a 1650-1800°C anneal, and diffusion doping does not work for SiC.  Finally, the epitaxial growth also occurs at high temperature (~1600°C).

A key fabrication challenge is to expose and etch these wafers and create the needed segmentation and junction isolation.  We are currently studying mask materials and designs, and performing preliminary etching tests on stock wafers.  The initial process flow and test devices will use a "mesa" structure, to be followed by implanted field control electrodes to achieve the needed packing density and fill factor.  Eventually the effort will have to confront the stringent requirements of fine segmentation by, for example, methods already being explored in the Silicon LGAD community [8]. We hope to benefit substantially from the knowledge base developed for power switching device fabrication.

Characterization of wafer and devices will include dark current, breakdown voltage and UV quantum efficiency.  Material property studies will include doping profiles (SIMS), structure (TEM), minority carrier lifetime, and defects (DLTS). In parallel we will establish a fast signal measurement capability to study charge collection and signal speed, before and after irradiation.

Based upon results from devices fabricated with the present set of epitaxial wafers received or ordered, we expect to iterate on the epi-layer thickness and doping levels in follow up orders.

**Conclusion**

Ultrafast timing is among the most important areas of current R&D in solid state charged particle tracking. It is transformative and compares in significance to the introduction of the strip detector in the 1970's, the strip readout ASIC, in the 1980's, the hybrid pixel detector, in the 1990's, and enhanced radiation resistant sensors and electronics in the 2000's. Fast timing will enable new physics and consequently, as stated, major experiments now plan to incorporate present generation silicon LGADs into special timing layers. Non-silicon solid state detectors have, so far, only found niche applications. We believe the combination of charge gain and large bandgap materials, here with SiC, and possibly others, could also be transformative. Gain overcomes the small signal problem, large bandgap means that the devices could be operated at high temperature, alleviating the need for massive cooling systems, and the growing industrial base for power devices means that scaled production may be possible. If the potential of this material is realized, both in performance and in availability, it could become the technological solution of choice for future HEP and NP fast timing systems.

**References**

[1] DOE Basic Research Needs Study on Instrumentation for HEP, 2020
https://science.osti.gov/hep/Community-Resources/Reports

[2] Hartmut F-W Sadrozinski *et al.,* "4D tracking with ultra-fast silicon detectors", 2018 *Rep. Prog. Phys.* **81** 026101 https://iopscience.iop.org/article/10.1088/1361-6633/aa94d3

[3] F. Nava *et al* , "Silicon carbide and its use as a radiation detector material", 2008 *Meas. Sci. Technol.* **19** 102001, • 2008 IOP Publishing Ltd
Measurement Science and Technology, Volume 19, Number 10

[4] https://en.wikipedia.org/wiki/Wide-bandgap_semiconductor

[5] Tao Yang *et al.,* "Time resolution of 4H SiC PIN and simulation of 4H SiC LGAD*", 38$^{th}$ RD50 Workshop.*

[6] Yuhang Tan *et al*.,"Time resolution simulation of 2D and 3D SiC detectors", *38$^{th}$ RD50 Workshop*

[7] T. Kimoto *et al.,* "Carrier lifetime and breakdown phenomena in SiC power device material", 2018 *J. Phys. D: Appl. Phys.* 51 363001

[8] G. Giacomini *et al*., "Fabrication and performance of AC-coupled LGADs", 2019 JINST **14**,
https://iopscience.iop.org/article/10.1088/1748-0221/14/09/P09004/pdf